\documentclass[12pt,date]{article}
\usepackage{amsmath}
\usepackage{graphicx}
\usepackage{amsfonts}

\newcommand{\p}{\partial}
\newcommand{\ds}{\displaystyle}
\newcommand{\beq}{\begin{eqnarray}}
\newcommand{\beqq}{\begin{eqnarray*}}
\newcommand{\eeq}{\end{eqnarray}}
\newcommand{\eeqq}{\end{eqnarray*}}
\newcommand{\n}{\mbox{\boldmath$n$}}
\begin{document}
\title{The search kinetics of a target inside the cell nucleus}
\author{G. Malherbe \thanks{D\'epartement de Math\'ematiques et de
Biologie, Ecole Normale Sup\'erieure, 46 rue d'Ulm 75005 Paris, France.} D. Holcman \thanks{Department of Mathematics, Weizmann Institute of Science,
Rehovot 76100, Israel. D\'epartement de Math\'ematiques et de Biologie, Ecole Normale Sup\'erieure, 46 rue d'Ulm 75005 Paris, France. D. H. is
supported by the program ``Chaire d'Excellence''.}}

\begin{abstract}
The mean time required by a transcription factor (TF) or an enzyme
to find a target in the nucleus is of prime importance for the
initialization of transcription, gene activation or the start of DNA
repair. We obtain new estimates for the mean search time when the TF
or enzyme, confined to the cell nucleus, can switch from a one
dimensional motion along the DNA and a free Brownian regime inside
the crowded nucleus. We give analytical expressions for the mean
time the particle stays bound to the DNA, $\tau_{DNA}$, and the mean
time it diffuses freely, $\tau_{free}$. Contrary to previous results
but in agreement with experimental data, we find a factor
$\tau_{DNA} \approx 3.7 \tau_{free}$ for the Lac-I TF. The formula
obtained for the time required to bind to a target site is found to
be coherent with observed data. We also conclude that a higher DNA
density leads to a more efficient search process.
\end{abstract}
\maketitle

\noindent{\textit{Introduction}} The search process for a target
promoter sequence by a transcription factor(TF) or for a double
strand break in the DNA by an enzyme such as Rac-A are fundamental
processes of cell activity and survival. In the first case, the
search process controls gene expression, while in the second, it
precedes DNA repair. In both cases timing is crucial as, for
example, unrepaired breaks are an obstacle for normal cell function
and can lead to mutations or apoptosis \cite{Minsky}.

The analysis of the mean time required for a TF to bind with a
promoter site originates from the early work of Berg-Von Hippel
\cite{Berg1,VonHippel,Berg}. They proposed a new but now well
accepted scenario to resolve the apparent paradox that this time
was, as experimentally observed, much faster than what it would be
if only free diffusion was involved. In this scenario, the TF can be
trapped by an unspecific potential energy and slide along the DNA
molecule. It then either finds its final target or detaches through
thermal noise and diffuses freely until it binds to another portion
of the DNA. This process iterates until the final site is reached.
Recent experiments have studied the kinetics of binding and
unbinding to the DNA using single particle tracking. In the case of
Lac-I, the time spent bound to the DNA represents about $87\%$
\cite{Elf} of the total search time.

The process of sliding along the DNA can be modeled as a sequence of
jumps between local potential wells resulting from the interaction
with the base pairs (bp). Approximating this motion by a reduced one
dimensional Brownian motion leads to large variance of the diffusion
coefficient \cite{Austin}, and today this approximation is thus
understood as a drastic simplification. A refined analysis was
developed in \cite{Slutsky}, for the mean number of bp scanned
during each binding to the DNA and time required to find the target
site by using some results on the motion in random environments
\cite{murthy}.

In this letter, we propose to revisit the computations of the search
time $\overline{T}_{S}$. We begin with the Berg-von-Hippel model
\cite{VonHippel}. In our model, a TF is confined in the nucleus,
which contains a set of DNA molecules. The interaction between the
TF and the DNA molecule is modeled by a potential well, obtained by
summing the specific and an unspecific potential\cite{Berg,Slutsky}.
The unspecific potential accounts for the interaction between the TF
and the DNA general structure, while the specific potential accounts
for the interaction between the TF and the DNA bp. Restricted by the
unspecific potential, the TF can slide and scan potential binding
sites along the DNA, until it detaches by thermal noise. Unbound,
the TF diffuses freely in the nucleus until it comes close enough to
the DNA where it can bind again.

To obtain an asymptotic estimate of the number of bp scanned per
binding $\overline{n}$ we generalize the computations of
\cite{Berg,Slutsky} by using the notion of random potential and the
solution of the mean first passage time equation \cite{SchussBook}.
Based on the narrow escape computations \cite{PNAS}, we estimate the
mean time $\overline{\tau}_{free}$ the TF spends in the nuclear
space before rebinding to a DNA molecule. Here the term "nucleus"
refers either to the nucleus of a eucaryotic cell or to the entire
bacteria for a procaryotic organism. The term "transcription factor"
(TF) refers either to a transcription factor or to a DNA binding
protein whose dynamical behavior can be modeled by the same general
assumptions (for example the Rac protein involved in DNA repair in
bacteria).

\noindent{\textit{General expression of the mean search time}}.
We recall the general expression of the mean search time,
$\overline{T}_{S}$, required by a TF to bind to its target
\cite{VonHippel}. We express it as a function of
$\overline{\tau}_{DNA}$ the time spent bound to the DNA,
$\overline{n}$ the mean number of bp scanned during this time,
$N_{bp}$ the total number of bp in the DNA and
$\overline{\tau}_{free}$ the time spent freely diffusing in the
nucleus. By conditioning on the number of bindings to the DNA, the
total search time is given by \cite{Berg1, Slutsky} \beq
\overline{T}_S\approx(\overline{\tau}_{DNA}+\overline{\tau}_{free})\frac{N_{bp}}{\overline{n}}
\label{final} \eeq Our goal is to obtain explicit formulas for
$\overline{n},\overline{\tau}_{DNA},\overline{\tau}_{free}$ as a
function of the geometry of nucleus, DNA distribution and other
physical parameters.

\noindent{\textit{Estimate of $\overline{\tau}_{DNA}$}} To estimate
$\overline{\tau}_{DNA}$, the average time a TF stays attached to the
DNA, we study the interaction potential with the DNA backbone. This
potential is mainly due to the charged phosphate groups of the DNA
backbone. We hence model it as a potential $V(r)=-\frac{k}{r}$ with
$r$ the distance to the DNA axis.

We account for the impenetrable condition between the TF and the DNA
molecule by defining a reflexive boundary condition at $r=R_{int}$,
the radius of the DNA double spiral. We consider that the TF is
freed when it reaches the position $r=R_{ext}$, which corresponds to
the maximal distance that allows bp discrimination. In practice, we
choose $R_{ext}=2R_{int}$. The mean time a TF starting at position
$r$ stays confined near the DNA molecule, $u(r)$, verifies
\cite{KarlinBook,SchussBook}: \beq
\Delta u(r)-\frac{\nabla V}{kT} \nabla u(r)=-\frac{1}{D}  \mbox{ for } R_{int}<r<R_{ext}\\
u(R_{ext})=0 \mbox{ and } \frac{\p u}{\p \n}(R_{int})=0 \nonumber
\label{Dynkin} \eeq where $\n$ is the normal vector to the reflexive
boundary. We solve this equation by direct integration and
approximate the expression obtained by a Laplace method: \beq
\overline{\tau}_{DNA}=u(r)\approx \frac{1}{D}
\frac{R_{int}(R_{int}-R_{ext})^{2}}{R_{ext}}\frac{(k_{B}T)^{2}}
{E_{ns}^{2}}e^{-\frac{E_{ns}}{k_{B}T}}. \label{TDNA}
 \eeq
where $E_{ns}=V(R_{ext})-V(R_{int})$ is the potential depth. For Lac
I and the parameters given in table \ref{table}
$\overline{\tau}_{DNA}\approx 5.7 ms$ which is compatible with
observed data ($\approx 5 ms$, \cite{Elf}).

\noindent{\textit{Mean number of sites scanned.} A TF whose motion
is restricted through interaction with the DNA is said to be
unspecifically bound. It then moves along the DNA driven by the
unspecific potential. We first estimate the average number of bp
visited for a constant specific potential. The number of sites
scanned during a time $\tau$, $n_{\tau}$, is equal to: \beq
n(\tau)=\frac{\max_{t\in [0;\tau]}(x(t))-\min_{t\in [0;\tau]}(x(t))}
{l_{bp}}, \eeq where $l_{bp}$ is the length of a bp and for a TF
whose position on the DNA is x(t) with x(0)=0. When the DNA molecule
is approximate by an infinite line, the distribution of the max (and
-min) is given by \cite{rednerbook} \beq \mathbb{P}\left(\max_{t\in
[0;\tau]}(x(t))\leq
x_{0}\right)=erf(\frac{x_{0}}{\sqrt{4D\tau}})\eeq where $erf(x)
=1-\frac{2}{\sqrt{\pi}} \int_{x}^{\infty} e^{-t^2}dt.$ This
distribution for the max of x during a time $\tau$ then allows us to
compute the mean value of the max for a given time $\tau$. The time
spent unspecifically bound is exponentially distributed for a
potential well deep before $k_BT$ \cite{SchussBook} and the mean
time, $\overline{\tau}_{DNA}$, is given in formula (\ref{TDNA}).
Thus the TF scans an average number of bp given by \beq
\overline{n}_{0}=\int_0^{\infty}
\overline{n}(\tau)e^{-\frac{\tau}{\overline{\tau}_{DNA}}}
d\left(\frac{\tau}{\overline{\tau}_{DNA}}\right)=
2\frac{\sqrt{D\overline{\tau}_{DNA}}}{l_{bp}}. \label{sigma0} \eeq
This leads to a 40\% increase compared to the mean square
displacement formula.

\noindent{\textit{Non constant specific potential.}} We consider a
more realistic model in which we estimate the probabilities and mean
time required to move one bp. We take into account the local
interaction between the TF and the DNA bp. Although such an approach
was considered in \cite{Slutsky}, our new estimate for the number of
bp visited $\overline{n}$ differs by a factor two compared to
\cite{Slutsky}.

\noindent{\textit{Number of bp scanned}.} The TF can move one bp to
the right (resp. to the left) with a probability $p_{i}$ (resp.
$q_{i}$) when bound to the DNA molecule. We let
$w_{i}=\frac{p_{i}}{q_{i}}$. Following the theory of random walk in
a random one dimensional potential \cite{murthy}, the average number
of steps $S_{0,N}$ needed by a TF to go from position 0 to N for the
first time, is given by: \beq
\overline{S}_{0,N}=N+\sum\limits_{k=0}^{N-1}w_{k}+
\sum\limits_{k=0}^{N-2}\sum\limits_{i=k+1}^{N-1}(1+w_{k})\prod\limits_{j=k+1}^{i}w_{j}.
\label{marche1D} \eeq If $\overline{u}$ denotes the average time
needed by the TF to move one bp, then the mean square displacement
during time a $\tau$ expressed in bp, $N_{\tau}$, is solution of
\beq \tau=\ds{\overline{u}\overline{S}_{0,N_{\tau}}}
\label{timestepsequ} \eeq

\noindent{\textit{Jump probabilities.}} The probability $p_{i}$ that
a TF at position $x(i)$, on bp $i$, moves to the right, satisfies
\cite{KarlinBook} \beq \label{eqproba} D\frac{\partial
^{2}p}{\partial x^2}-\frac{D}{k_BT}\frac{\partial
V}{\partial x}\frac{\partial p}{\partial x}=0 \\
p(x(i-1))=0 \hbox{ and } p(x(i+1))=1. \nonumber \eeq For a piecewise
constant potential $V$, equal to $E_i$ near bp $i$, we solve
equation (\ref{eqproba}) and: \beq\label{wi} w_{i}=
\frac{p_{i}}{q_{i}}=\frac{p_{i}}{1-p_{i}}=\ds{\frac{e^{\frac{E(i-1)}{k_BT}}+e^{\frac{E(i)}{k_BT}}}{e^{\frac{E(i+1)}{k_BT}}+e^{\frac{E(i)}{k_BT}}}}.
\eeq

\noindent{\textit{Average time required to move one bp.}} To
evaluate expression (\ref{timestepsequ}), we estimate the mean time
$\overline{u}$ required by a TF to move one bp. It is the solution
of Dynkin's equation given in \ref{Dynkin} with the absorbing
conditions $u(x(i-1))=u(x(i+1))=0$ We explicitly solve this equation
and obtain the average time $u_{i}=u(x(i))$ to move one step to the
left or to the right and for a piecewise potential:\beq
u_{i}=\frac{l_{bp}^{2}}{2D}\left(1+\frac{3}{2}\frac{e^{\frac{E_{i+1}-E_{i}}
{kT}}e^{\frac{E_{i-1}-E_{i}}{kT}}-1}{e^{\frac{E_{i+1}-E_{i}}{kT}}+e^{\frac{E_{i-1}-E_{i}}{kT}}+2}\right)
\label{meantimeonestep}, \eeq where $l_{bp}$ is the average length
of a bp.

\noindent{\textit{Number of potential binding sites scanned}} We
denote by $i$ the position of the TF's beginning n the number of bp
interacting with the TF. The position weight matrix model
\cite{Stromo, Takeda} has already been shown to be equivalent to a
normal distribution of $E_{i}$, the specific energy of a given site
i (\cite{Slutsky}). In addition the specific energies for sites
starting at positions i and j can be correlated. For $|i-j|\geq n$
the specific energies are independent. For $|i-j|<n$ there are
$n-|i-j|$ bp contributing to both energies that induce a correlation
between the energies for the sites i and j. One can further show by
taking linear combinations $\alpha E_{i}+\beta E_{i+1}$ that
$(E_{i},E_{i+1})$ follows a bivariate normal law.

We can then estimate expression (\ref{marche1D}) by neglecting the
two terms of order $N$ in front of the term of order $N^2$. With
$\overline{x}_\tau=N_{\tau}l_{bp}$ the mean square displacement:

\beq\overline{S}_{0,N_{\tau}}\simeq
\left(\frac{\overline{x}_\tau}{l_{bp}}^{2}\right)\mathbb{E}_{(E_{k})_{k\in
\mathbb{N}}}\left(\frac{e^{\frac{E(j+1)}{kT}}+e^{\frac{E(j)}{kT}}}{e^{\frac{E(i+1)}{kT}}+e^{\frac{E(i)}{kT}}}\right)
\label{meanstepsbeforeaverage}\eeq for couples such that $\mid
j-i\mid >n$. We can then average over the different energy levels
and find an estimate with a Laplace method. Similarly we estimate
$\overline{u}$ by averaging (\ref{meantimeonestep}) over the energy
levels. We then obtain $\overline{x}_\tau$ with equation
(\ref{timestepsequ}). By considering the mean square displacement is
proportional to the average number of bp scanned and with formula
(\ref{sigma0}), we obtain $\overline{n}$ the mean number of bp
visited during a typical one dimensional walk, \beq
\overline{n}\simeq2\sqrt{\frac{D\overline{\tau}_{DNA}
e^{-\frac{\sigma^2}{2(kT)^{2}}}e^{-\frac{\sigma^2(1+\rho)}{4(kT)^{2}}}
\sqrt{1+\frac{\sigma^{2}(1-\rho)}{2(kT)^{2}}}}
{l_{pb}^{2}\left(1+\frac{3e^{\frac{3\sigma^{2}(1-\rho)}{4(kT)^{2}}}}{4\sqrt{1+\frac{\sigma^{2}(1-\rho)}
{2(kT)^{2}}}}-\frac{3}{4\sqrt{1+\frac{\sigma^{2}(1-\rho)}{(kT)^{2}}}}\right)}},
\label{nbarre} \eeq with $\sigma=\sqrt{\mathbb{E}(E_i^2)}$ the
variance and $\rho=\frac{\mathbb{E}(E_{i}E_{i+1})}{\sigma^{2}}$ the
correlation factor. In figure \ref{effectsigma}, we show how
$\overline{n}$ depends on $\sigma$ and $\rho$. For large $\sigma$,
we approximate by: \beq \overline{n}= \ds{
\overline{n}_{0}\sqrt{\frac{4}{3}}e^{-\frac{\sigma^{2}}{2(kT)^{2}}}
e^{-\frac{\sigma^{2}(1-\rho)}{4(kT)^{2}}} } \nonumber \eeq where
$\overline{n}_{0}$ is given in equation (\ref{sigma0}). We find
$\overline{n}=75$ sites visited during a typical one dimensional
walk of $5ms$ with the data in table \ref{table}. This can be
compared with the experimental value of $\approx 85$ \cite{Elf}.

\begin{figure}
\includegraphics[width=1.5in]{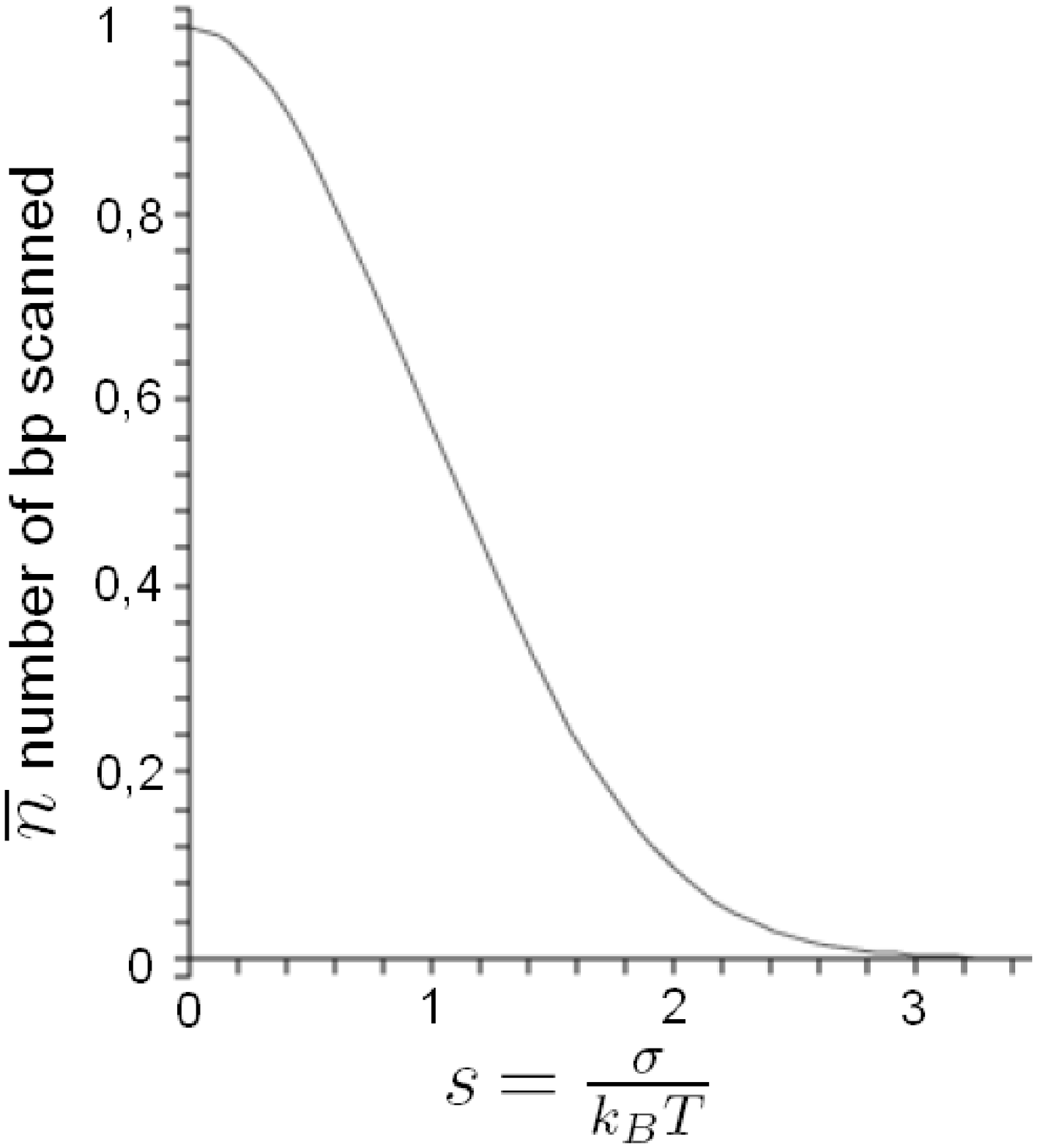}
\includegraphics[width=1.5in]{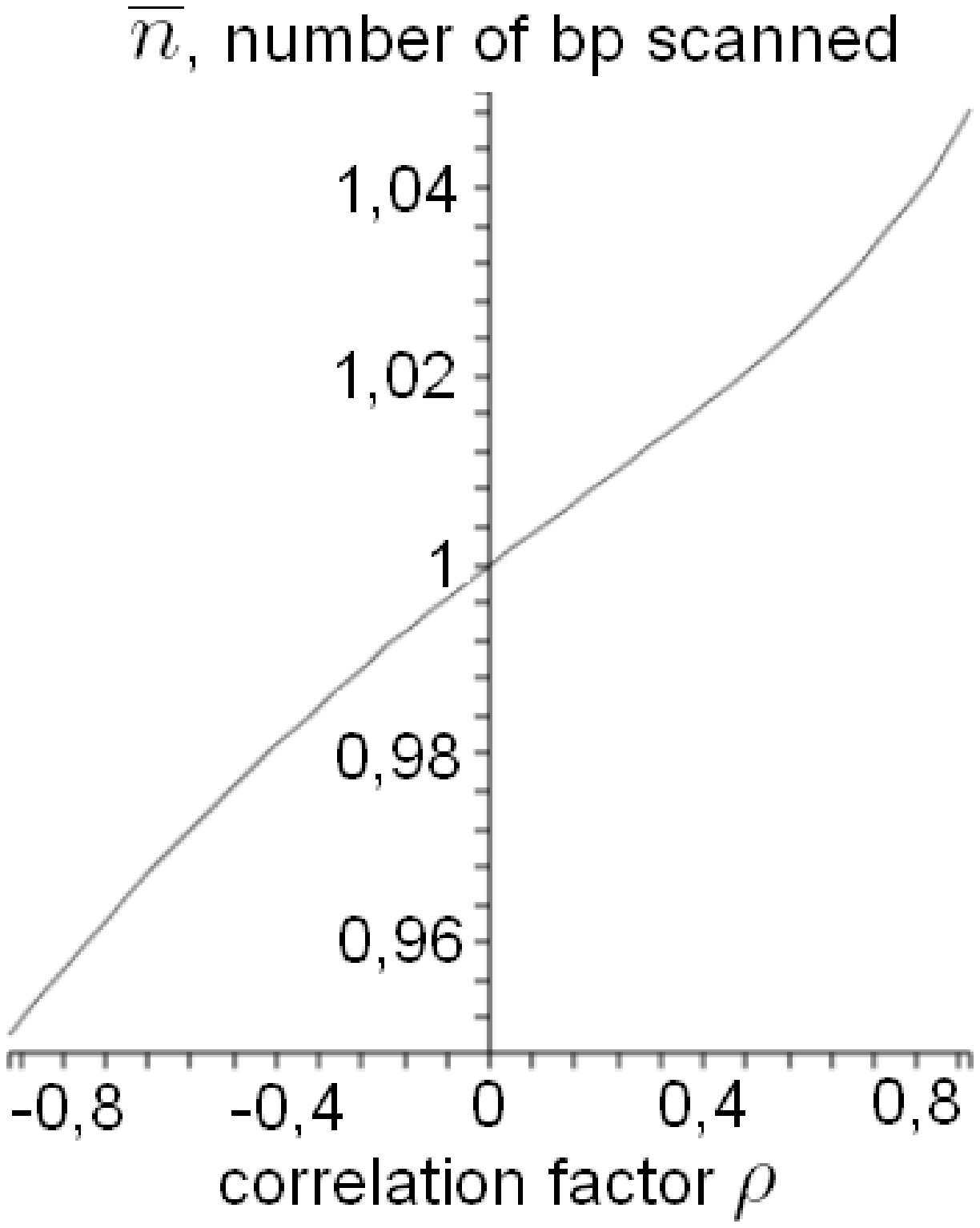}
 \caption{{Left:} Mean number of sites visited as a function of
$s=\frac{\sigma}{k_{B}T}$ for no correlation ($\rho=0$) expressed in
multiples of $\overline{n}_{0}$. Right: Number of sites visited for
$\sigma=k_{b}T$ as a function of the correlation factor $\rho$ in
multiples of the value for $\rho=0$. A positive correlation
$\rho>0$, is associated with a lesser apparent roughness of the
specific potential and to a more sites scanned. With $\rho<0$, low
energy sites have a tendency of being flanked by high energy sites
which leads to a greater number of local minima of the specific
potential and to less sites visited.} \label{effectsigma}
\end{figure}
\noindent{\textit{Free diffusion time}.} We now estimate
$\overline{\tau}_{free}$, the mean time a TF freely diffuses in the
nucleus between two consecutive DNA bindings. As stated at the end
of the introduction the term "nucleus" refers either to the nucleus
of an eucaryotic organism (modeled as a sphere of radius R) or the
entire bacteria for a procaryotic organism (modeled as a cylinder of
radius R for E Coli). We consider the DNA is organized (Fig.
\ref{absorbtiontime}) on a square lattice of $N_{st}$ parallel
cylindrical strands of diameter
$2\epsilon=2R\sqrt{\frac{\rho_{DNA}}{N_{st}}} \approx 30nm$ where
$\rho_{DNA}$ is the ratio of absorbing DNA volume to the total
nuclear volume. These strands account for the DNA structure below
the 30nm fibers. We consider here such organization, which has been
observed in some bacteria after cell irradiation \cite{Minsky}.

\begin{figure}
\includegraphics[width=1.5in]{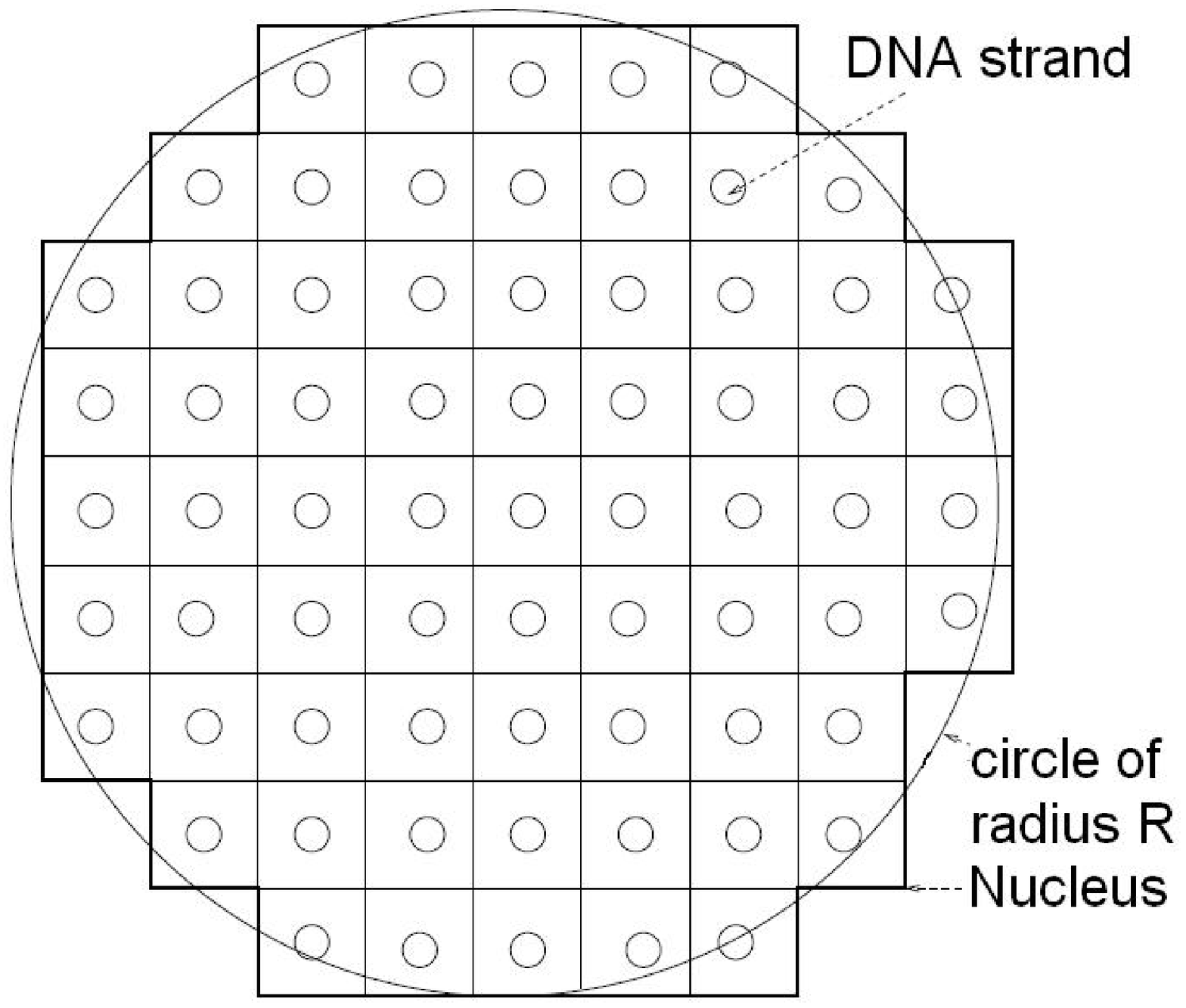}
\includegraphics[width=1.5in]{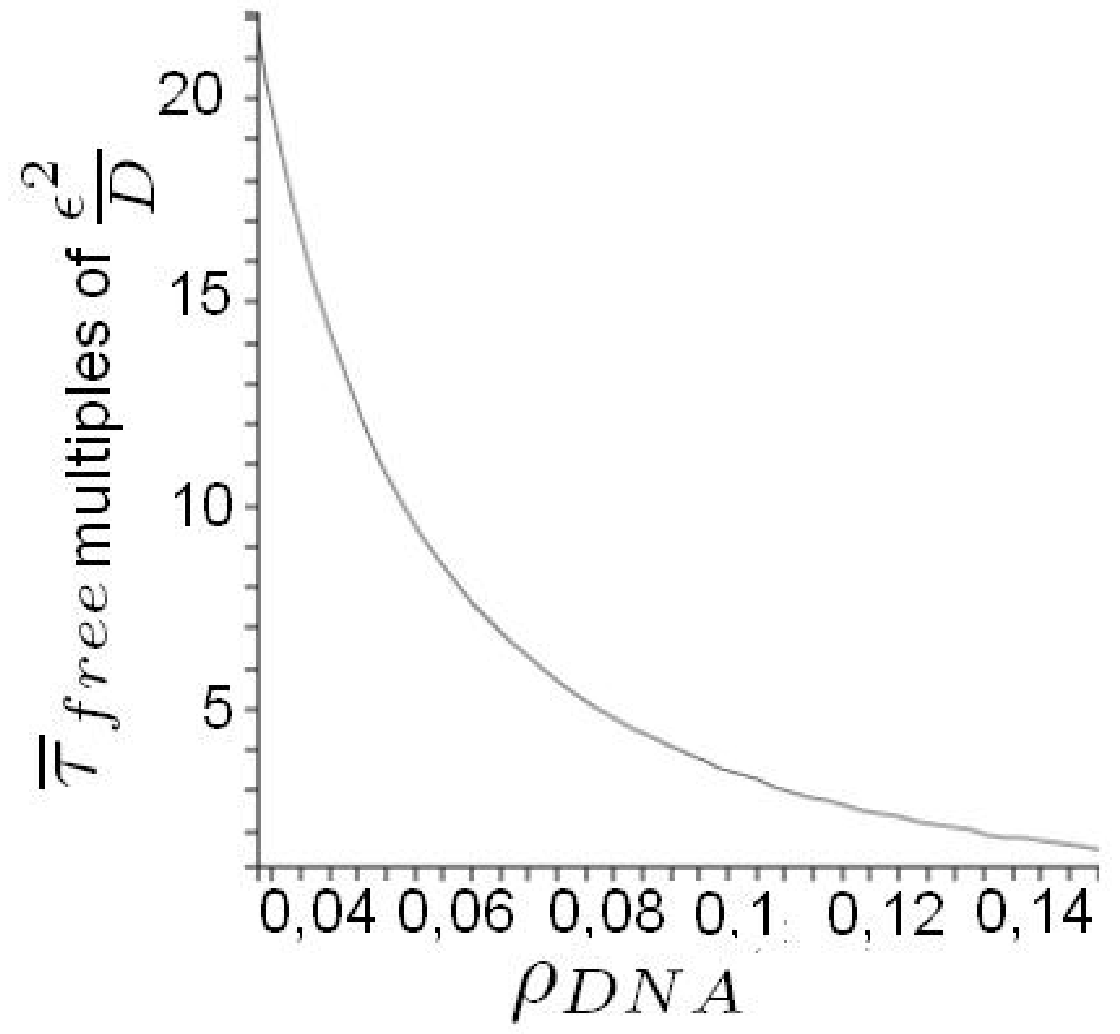}
 \caption{{Left:} Schematic of a two dimensional section of the nucleus
 perpendicular to the DNA organized on a square lattice. Each DNA strand is a compacted in a 30 nm
  fiber. We approximate the nucleus by a collection of boxes. {Right:} Free diffusion
 time $\overline{\tau}_{free}$ in multiples of $\frac{\epsilon^{2}}{D}$ plotted as a function of $\rho_{DNA}$
 the DNA density. High DNA density leads to a faster DNA search.
$\frac{\epsilon^{2}}{D}\approx 0.1 ms.$ for the parameters of table
\ref{table}} \label{absorbtiontime}
\end{figure}

For parallel DNA strands, we can, by symmetry, consider only a
single two dimensional square (Fig. \ref{absorbtiontime}). The TF is
absorbed at the external radius $\epsilon$, and is considered to be
reflected on the square boundary as it enters a symmetrical and
identical square. In cylindrical coordinates the mean time before
absorbtion, $u(r,\theta)$, when starting at position $(r,\theta)$
verifies \cite{SchussBook}: \beq \left\{\begin{array}{l}
\ds{\frac{1}{r}\frac{\partial}{\partial r}\left( r\frac{\partial
u}{\partial r}\right)+\frac{1}{r^{2}}\frac{\partial^{2}u}{\partial
\theta^{2}}=-\frac{1}{D}}\\
\ds{\frac{\p u}{\p \n}}=0 \mbox{ on the square of size }\sqrt{\frac{\pi
R^{2}}{N_{st}}}
\\\ds{u(r, \theta)}=0 \mbox{ for } r=\epsilon,\mbox{ for } \theta
\in [0,2\pi],
\end{array}\right.,
\label{dynkin3D} \eeq where $\n$ is the normal vector to the square
boundary. The solution can be expressed in the form \beq
u=u_{0}+A\ln\left(\frac{r}{\epsilon}\right)
+\sum_{n=1}^{\infty}(A_{n}\left(\frac{r}{\epsilon}\right)^{n}
+B_{n}\left(\frac{r}{\epsilon}\right)^{-n})\cos(n\theta)
\label{fourier3D}\nonumber\eeq where $A,A_{n}$ and $B_{n}$ are
constants to be determined and
$u_{0}=-\frac{r^{2}-\epsilon^{2}}{4D}$. The absorbing boundary
condition at $r=\epsilon$ requires $A_{n}=-B_{n}$. Moreover, by
symmetry, only $A$, $A_{4n}$ and $B_{4n}$ are non null.
$\overline{\tau}_{free}$ is the average of u over a uniform initial
distribution. We need to estimate the coefficient $A$ and the other
remaining terms since they have a contribution due to the effect of
the corners. To find the coefficients, we use the reflective
boundary condition. We let: \beq
B_{0}&=& A\frac{8D\rho_{DNA}}{\pi \epsilon^{2}} \\
B_{n}&=& \ds{4n A_{4n}\frac{8D\rho_{DNA}}{\pi \epsilon^{2}
}}\left(\frac{\pi}{4\rho_{DNA}}\right)^{2n}, \hbox{ for } n>0
\label{defnormalizedcoefs} \eeq by neglecting
$\left(\frac{\epsilon}{r}\right)^{4n}$ in front of
$\left(\frac{r}{\epsilon}\right)^{4n}$ and for
$\theta\in[0;\frac{\pi}{4}]$: \beq
0=-\frac{\tan(\theta)}{\cos^{2}\theta}+B_{0}\tan(\theta)+\sum\limits_{n=1}\limits^{\infty}
B_{n}\frac{\sin((4n+1)\theta)}{\cos^{4n+1}\theta} \label{seriecos}
\eeq By expanding in variable $\xi=\tan\theta$, we obtain a power
series and identify the terms of same degree. We can then
numerically solve the infinite system of algebraic equations by
truncating at a certain rank. Finally, by reporting into the
expression of $u$ and after averaging over a uniform initial
position: \beq
\overline{\tau}_{free}=\frac{\epsilon^{2}}{D\rho_{DNA}}
\left(0.3\ln(\rho_{DNA})-0.41+0.55\rho_{DNA}\right) \label{freetime}
\eeq where $\rho_{DNA}$ is the the ratio of the absorbing DNA to the
total nuclear volume. In figure \ref{absorbtiontime}, we plot
$\overline{\tau}_{free}$ as a function of $\rho_{DNA}$, the ratio of
the absorbing DNA to the total nuclear volume.

\begin{table}
\caption{Numerical parameters and results for for LacI}
\begin{tabular}{l l l l}
  \hline
 $D$ & LacI diffusion constant& $3\mu m^{2}.s^{-1}$
&\cite{Elf}\\
$N_{bp}$&Number of bp in E Coli&$4.8*10^6$& \\
$l_{bp}$ & Average length of a bp& $0.34nm$& \\
$R$& Radius of E Coli& $0.4\mu m$& CCDB database \\
$L$& Length of E Coli& $2\mu m$& CCDB database \\
$E_{ns}$& Non specific energy &
-16kT&\cite{Gerland} \\
$\sigma$ &  Spec. energy roughness &2kT&Regtrans base\\
$\rho$ & Correlation factor&$+2\%$&Regtrans base\\
$2\epsilon$ & Diameter of DNA fiber&$30nm$& \\
$R_{int}$&DNA double helix radius&$1nm$ & \\
$R_{ext}$&Potential external radius&$2nm$&
\end{tabular}
\begin{tabular}{l l l}
\hline
$\overline{\tau}_{free}$& Average time spent freely diffusing&1.5 ms \\
 $\overline{\tau}_{DNA}$& Average time spent unspecifically bound&5.7 ms\\
 $\overline{n}$& Average number of bp scanned &75\\
$\overline{T}_S$& Average time needed to find the target site&7min40\\
 \hline
\end{tabular}
\label{table}
\end{table}

It is interesting to note that, when multiplying $N_{bp}$ and
$\rho_{DNA}$ by a factor k (this increases the DNA density by a
factor k and keeps the nucleus volume constant), the global search
time given in \ref{final} is multiplied by a factor strictly smaller
than k while searching through k times more information. We conclude
that a higher DNA density leads to a more efficient search process.

Using formula (\ref{nbarre}),(\ref{TDNA}) and (\ref{freetime}) and
the data given for E. Coli in table \ref{table}, we obtain
$\overline{\tau}_{free}=1.5 ms$, $\overline{\tau}_{DNA}=5.7 ms$,
$\overline{n}=75$ and an average search time of $\bar{T}_{S}=7 min
48 s$, which is compatible with observed data \cite{Elf}.  Our
conclusions do not rely on the assumption that
$\overline{\tau}_{free}=\overline{\tau}_{DNA}$ as we obtain two
independent expressions for $\overline{\tau}_{free}$ and
$\overline{\tau}_{DNA}$. Moreover, we find that a TF stays bound to
the DNA molecule for roughly 80 \% of the total search time. This
agrees with the experimental data published in \cite{Elf}, where the
TF is bound around 87 \% of the time to the DNA molecule. It would
be an interesting problem to extend our method to a more general DNA
distribution.

{\bf Acknowledgements:} D. H. is supported by the program ``Chaire
d'Excellence''.


\end{document}